\documentclass[5p,twocolumn,times,number]{elsarticle}

\usepackage{graphicx}
\usepackage{amsmath}   
\usepackage{lineno}

\journal{Nuclear Instruments and Methods A}

\begin{document}

\begin{frontmatter}

\title{Barrel time-of-flight detector for the PANDA experiment at FAIR}

\author[add1,add2]{L.~Gruber\corref{cor}}
\ead{lukas.gruber@oeaw.ac.at}
\author[add1]{S.~E.~Brunner}
\author[add1]{J.~Marton}
\author[add2]{H.~Orth}
\author{K.~Suzuki$^{\mathrm{a}}$ on behalf of the PANDA TOF Group}

\cortext[cor]{Corresponding author}

\address[add1]{Stefan Meyer Institute for Subatomic Physics, Austrian Academy of Sciences, Vienna, Austria}
\address[add2]{GSI Helmholtz Centre for Heavy Ion Research, Darmstadt, Germany}


\begin{abstract}
The barrel time-of-flight detector for the PANDA experiment at FAIR is foreseen as a Scintillator Tile (SciTil) Hodoscope based on several thousand small plastic scintillator tiles read-out with directly attached Silicon Photomultipliers (SiPMs). The main tasks of the system are an accurate determination of the time origin of particle tracks to avoid event mixing at high collission rates, relative time-of-flight measurements as well as particle identification in the low momentum regime. The main requirements are the use of a minimum material amount and a time resolution of $\sigma < 100\,\mathrm{ps}$. We have performed extensive optimization studies and prototype tests to prove the feasibility of the SciTil design and finalize the R\&D phase. In a 2.7\,GeV/c proton beam at Forschungszentrum J\"ulich a time resolution of about 80\,ps has been achieved using SiPMs from KETEK and Hamamatsu with an active area of $3\times3\,\mathrm{mm^2}$. Employing the Digital Photon Counter from Philips a time resolution of about 30\,ps has been reached.
\end{abstract}


\end{frontmatter}


\section{Introduction}

The Facility for Antiproton and Ion Research (FAIR)~\cite{fair} is currently being constructed as an expansion of the GSI Helmholtz Centre in Darmstadt, Germany. PANDA~\cite{panda} is one of the four major experiments at FAIR and will be located at the High Energy Storage Ring (HESR). Cooled beams of antiprotons with high luminosity up to $L = 2\times10^{32}\,\mathrm{cm^{-2}s^{-1}}$ and high resolution ($\delta p/p < 10^{-4}$) in the momentum range of 1.5\,GeV/c to 15\,GeV/c will be used for PANDA to perform high precision experiments in the strange and charm quark sector.

The PANDA detector is divided into two magnetic spectrometers, allowing almost 4$\pi$ angular coverage. Charged particle identification within the target spectrometer is done by Cherenkov detectors based on the DIRC principle and a fast timing detector in combination with the tracking system.    

\section{Barrel time-of-flight detector}

The design of the PANDA barrel time-of-flight (TOF) detector is motivated by physical as well as technical considerations. The main tasks of the system are particle identification for slow charged particles ($< 700\,\mathrm{MeV/c}$) below the momentum threshold of the DIRC detector, precise determination of the time origin of particle tracks to disentagle subsequent events at high collision rates of N$_{\mathrm{avg}} = 20\,\mathrm{MHz}$, relative time-of-flight measurements together with the forward TOF system and charge discrimination as well as photon conversion detection in front of the electromagnetic calorimeter (EMC). The major requirements are a minimum use of material less than 2\% of a radiation length, a time resolution of $\sigma < 100\,\mathrm{ps}$ and a large solid angle coverage of 22$^{\circ} \leq \theta \leq 140^{\circ}$.

The barrel TOF detector is located in between the DIRC detector and the EMC at a radial distance of about 50\,cm from the beam pipe. The extremely limited space of only 2\,cm in radial direction demands a lightweight and compact detector design as it has been proposed in Ref.~\cite{scitil}. The suggested layout foresees a Scintillator Tile (SciTil) Hodoscope composed of 5760 small plastic scintillator tiles with sizes of about $30\times30\times5\,\mathrm{mm^{3}}$, covering a total area of about 5.2\,m$^{2}$. The tiles are attached to Silicon Photomultipliers (SiPMs). Plastic scintillators provide fast response and high light yield while SiPMs offer good timing, high photon detection efficiency, compactness and insensitivity to the magnetic field of about 2\,T within the PANDA target spectrometer.

\section{Detector performance}

In order to fulfill the requirements for the SciTil detector, extensive optimization studies have been performed by Monte Carlo simulation and experiment. Results of these studies are reported in Refs.~\cite{optimization, diss}. To finally prove the feasibility of reaching a time resolution of $\sigma < 100\,\mathrm{ps}$ with the proposed detector, a SciTil prototype based on the optimization studies has been tested in a 2.7 GeV/c proton beam at Forschungszentrum J\"ulich. The beam intensity was adjusted to 10$^{4}$-$10^{5}$\,Hz and the beam size was approximately $5\,\mathrm{cm}\times5\,\mathrm{cm}$. The detector consisted of two plastic scintillator tiles read-out with SiPMs from two opposing sides mounted in a light-tight aluminum box. The SiPMs were coupled to the scintillators using BC-630 optical grease to maximize the light collection. We considered plastic scintillators from Eljen and analog SiPMs with $3\times3\,\mathrm{mm^2}$ sensitive area from KETEK and Hamamatsu, respectively, as well as the Digital Photon Counter (DPC) from Philips with an active area of about $32\times32\,\mathrm{mm^2}$ divided into $4\times4$ individual die sensors. The analog SiPM signals were amplified using Amp-0604 preamplifiers from Photonique. Finally, the waveforms were digitized with a CAEN V1742 Waveform Digitizer (5 GS/s) and stored for offline data analysis. A time walk correction using pulse height information was applied. In case of the DPC the electronics are already integrated inside the detector, providing digital energy and time information. 

\begin{figure}
  \centering
  \includegraphics[width=0.93\linewidth]{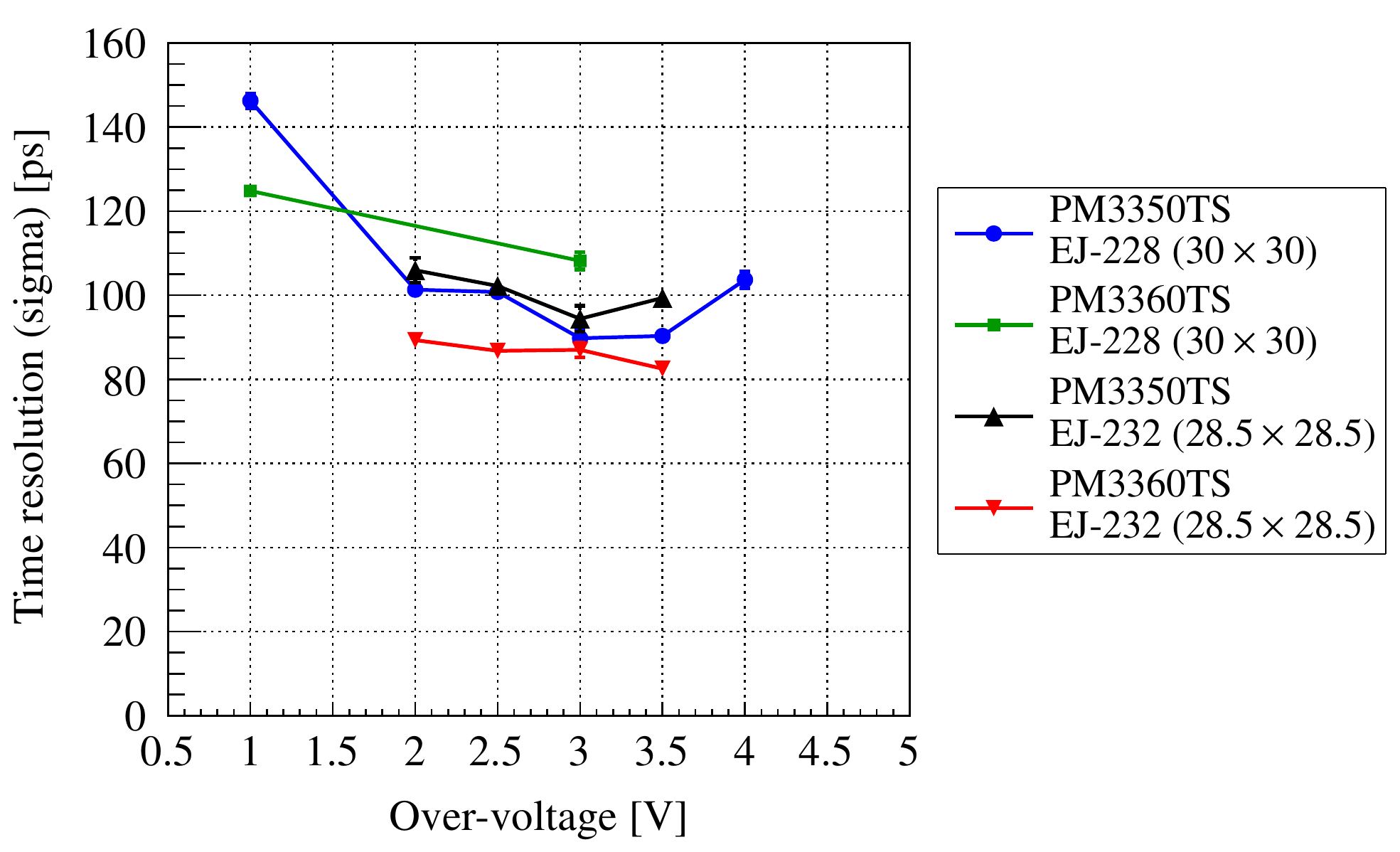}
  \caption{Time resolution of various plastic scintillator tiles with 5\,mm thickness as a function of over-voltage read-out with two KETEK SiPMs of different type. The size of the scintillator is indicated.}
  \label{fig:ResultsKetek}
\end{figure}

In the test beam experiment we observed that the scintillator tiles read-out with KETEK SiPMs show in general slightly better time resolution than the tiles attached to sensors from Hamamatsu. Fig.~\ref{fig:ResultsKetek} shows the obtained results on the time resolution using different types of SiPMs from KETEK. Except for the PM3360TS type in combination with an EJ-228 scintillator all configurations show a time resolution $\sigma < 100\,\mathrm{ps}$. The best value that could be achieved is $\sigma = 82.5 \pm 1.7\,\mathrm{ps}$ with PM3360TS sensors attached to an EJ-232 plastic scintillator. One can realize that the tile time resolution strongly depends on the over-voltage of the photodetectors. Increasing the bias voltage leads to improved time resolution but at the same time increases the influence of dark-noise, cross-talk and after-pulses, which at some point deteriorate the timing performance. The same trend could be observed when employing Hamamatsu photodetectors. In this case a time resolution of $\sigma = 96.3 \pm 0.9\,\mathrm{ps}$ was reached with the S12572-025C type attached to an EJ-228 scintillator. Using an EJ-232 material showed a nearly identical result.

\begin{figure}
  \centering
  \includegraphics[width=0.655\linewidth]{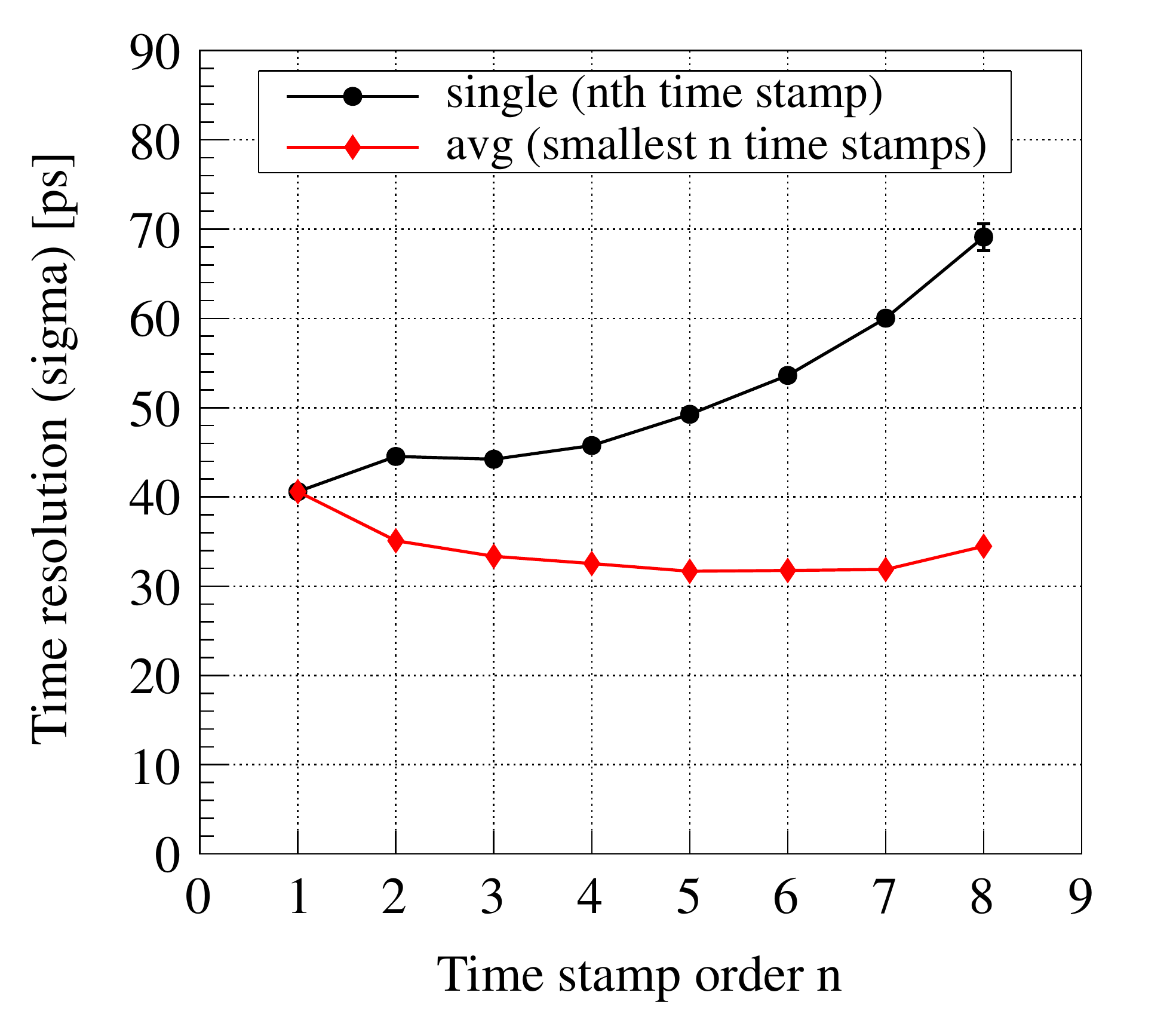}
  \caption{Tile time resolution using the Philips DPC attached to an EJ-228 scintillator with a size of $30\times30\times5\,\mathrm{mm^{3}}$. The time resolution was determined by using single die time stamps or an average of multiple time stamps.}
  \label{fig:ResultsDPC}
\end{figure}

Employing the large-area DPC allowed to cover a larger fraction of the scintillator surface. This led to an increased number of detected photons and a multitude of available trigger time stamps (one per die). The time resolution of the tile can either be estimated by using single time stamps or an average of multiple die time stamps. As illustrated in Fig.~\ref{fig:ResultsDPC}, a multiple time stamp approach provides a more accurate estimation and leads to improved time resolution, as expected. To perform this analysis the individual die time stamps have been sorted in ascending order. The best result of $\sigma = 31.7 \pm 0.1\,\mathrm{ps}$ could be achieved by using an average of the first five acquired time stamps. Taking into account higher time stamp orders, i.e. late photons, does not provide valuable input for this method. In case of the single time stamp approach the first order, i.e. the first detected photon, provided the best time precision, as expected from photon counting statistics~\cite{diss}.      

\section{Conclusion and outlook}
Experimental results have demonstrated that the envisaged time resolution of $\sigma < 100\,\mathrm{ps}$ for the SciTil detector is given with the proposed detector layout. Highest time resolution could be achieved with the large-area DPC by making use of multiple time stamps, though conventional $3\times3\,\mathrm{mm^2}$ SiPMs also met the requirements. Current effort is placed on the R\&D of read-out electronics and further optimization of the scintillator geometry. We are currently investigating bar-shaped scintillator tiles up to a size of $90\times30\times5\,\mathrm{mm^{3}}$ read-out from two sides by 3-4 SiPMs connected in series, as reported in Ref.~\cite{meg}. With such a design the time resolution can further be improved.

\section*{Acknowledgments}

This work was partly supported by the EU Project HadronPhysics3 (project 283286). The authors also thank the team of Philips Digital Photon Counting.


\end{document}